\documentclass[fleqn,usenatbib]{mnras}

\usepackage{newtxtext,newtxmath}

\usepackage[T1]{fontenc}
\usepackage{ae,aecompl}

\usepackage{graphicx}	
\usepackage{amsmath}	

\title[Star/Galaxy/AGN Classification]{Classifying Stars, Galaxies and AGN in  CLAUDS+HSC-SSP Using Gradient Boosted Decision Trees}

\author[A. Golob et al.]{
Anneya Golob$^{1}$\thanks{E-mail: \tt anneyagolob@gmail.com},
Marcin Sawicki$^{1}$\thanks{E-mail: {\tt marcin.sawicki@smu.ca}}\thanks{Canada Research Chair},
Andy~D.~Goulding$^{2}$,
Jean~Coupon$^{3}$
\newauthor
\\
$^{1}$Institute for Computational Astrophysics and 
Department of Astronomy and Physics, Saint Mary's University, 923 Robie Street, \\ Halifax, Nova Scotia B3H 3C3, Canada\\
$^{2}$Department of Astrophysical Sciences, Princeton University, Princeton, NJ 08540, USA\\
$^{3}$Astronomy Department, University of Geneva, Chemin d'Ecogia 16, CH-1290 Versoix, Switzerland\\
}

\date{Accepted XXX. Received YYY; in original form ZZZ}

\pubyear{2017}

\begin{document}
\label{firstpage}
\pagerange{\pageref{firstpage}--\pageref{lastpage}}
\maketitle

\begin{abstract}
Classifying catalog objects as stars, galaxies, or AGN is a crucial part of any statistical study of galaxies. We describe our pipeline for binary (star/galaxy) and multiclass (star/galaxy/Type I AGN/Type II AGN) classification developed for the very deep CLAUDS+HSC-SSP $u^*grizy$ dataset. Our method uses the {\tt XGBoost} implementation of Gradient Boosted Trees (GBT) to train ensembles of models which take photometry, colours, maximum surface brightnesses, and effective radii from all available bands as input, and output the probability that an object belongs to each of the classes under consideration. At $i_\mathrm{AB}<25$ our binary star/galaxy model has AUC=0.9974 and at the threshold that maximizes our sample's weighted F1 score, selects a sample of galaxies with 99.7\% purity and 99.8\% completeness. We test the model's ability to generalize to objects fainter than those seen during training and find that extrapolation of $\sim1-2$ magnitudes is reasonable for most applications provided that the galaxies in the training sample are representative of the range of redshifts and colours of the galaxies in the target sample. We also perform an exploratory analysis of the method's ability to identify AGN using a small x-ray selected sample and find that it holds promise for classifying type I AGN, although it performs less well for type II AGN.  Our results demonstrate that GBTs provide a flexible, robust and efficient method for performing classification of catalog objects in large astronomical imaging surveys. 
\end{abstract}



\section{Introduction}\label{sec:intro}

Classifying each of the objects detected in an astronomical survey as an enormous, distant galaxy or a relatively tiny star residing within our own Galaxy is a notoriously tricky task, but one that is fundamental to all manner of astrophysical analyses. If the objects in question are well-resolved, they can easily be distinguished in images as either point-sources (stars or, maybe, AGN) or extended objects (very likely galaxies). As such, measurements of morphology are very useful for making this classification (very bright galaxies are most generally nearby so that their projected extent will be quite obvious compared to stars of similar brightness) but offer little to no help in dealing with faint objects (faint galaxies are typically unresolved in ground-based images).  Similarly, if the wavelength coverage and spectral resolution of observations for the objects in question are favourable, the distinction can be made with high confidence: faraway galaxies will be redshifted and the range of realistic stellar spectra is well-known. As astronomy moves into the era of big data and discovery is driven by large-area imaging surveys pushing deeper and deeper, the overwhelming majority of identified sources that must be classified are unresolved, faint, and observed at only a handful of wavelengths. This is the challenge we face with our CLAUDS+HSC-SSP dataset \citep{clauds, hscssp} that probes the Universe to an unprecedented combination of area and depth ($Ugrizy$ imaging to $\sim$27~AB over $\sim$20~deg$^2$ -- see Sec.~\ref{sec:data}).  

To segregate galaxies from stars, current large astronomical surveys use techniques such as photometric template fitting (e.g. \citealt{lephare}), specific colour-selection criteria (e.g. \citealt{bzk, AO2013}),  the comparison of objects' maximum surface brightness and total magnitude (e.g. \citealt{acsCat}, \citealt{bardeau}), or combinations of these methods (e.g. \citealt{laigle2017}, \citealt{thibaud}). The poorly-characterized objects found in astronomical datasets, combined with the high classification accuracy required by the science cases they'll be used to probe (e.g., \citealt{Soumagnac2015}) push the limits of the techniques currently used in practical survey applications.  For this reason,  we decided to develop a technique that combines the available photometric and morphological information when faced with the star-galaxy classification challenge in our CLAUDS+HSC-SSP dataset. 

In recent years, the field of data science has blossomed, providing a plethora of open-source implementations of cutting-edge algorithms to deal with difficult classification problems. Labeling objects as stars or galaxies is, in theory, a prototypical supervised learning problem, but astronomical datasets often have noisy labels and training samples with high-confidence labels are rare and frequently biased. Despite these challenges, the astronomical community has worked to apply machine learning methods to the star/galaxy classification problem such as neural networks (e.g. \citealt{sextractor}, \citealt{sgspNN1992}, \citealt{Soumagnac2015}), convolutional neural networks (e.g., \citealt{sgspConvNN2017}), support vector machines (e.g., \citealt{sgspSVMwise2015}), decision trees (e.g. \citealt{sgspDecisionTrees2011, Costa-Duarte2019}, \citealt{sgspAdaBoost2015}), random forests (e.g., \citealt{Bai2018}), 
and ensemble methods (e.g. \citealt{sgspHybrid2015}).\\

Comparing the performance of classifiers described in the literature which have been applied to different samples is nearly impossible because the outcome is highly sensitive to the features used by the model and the characteristics of the training and test data. Various works have sought to make fair comparisons of different algorithms on specific datasets (e.g. \citealt{Machado2016}, \citealt{fadely2012}, \citealt{sgspAdaBoost2015}), but these investigations find that there is no best classification algorithm for star-galaxy classification. The best method in a given situation depends on the requirements of the science case in question, the specific characteristics of the data and underlying population, and the test sample available.

\begin{figure}
\begin{centering}
\includegraphics[width=0.48\textwidth]{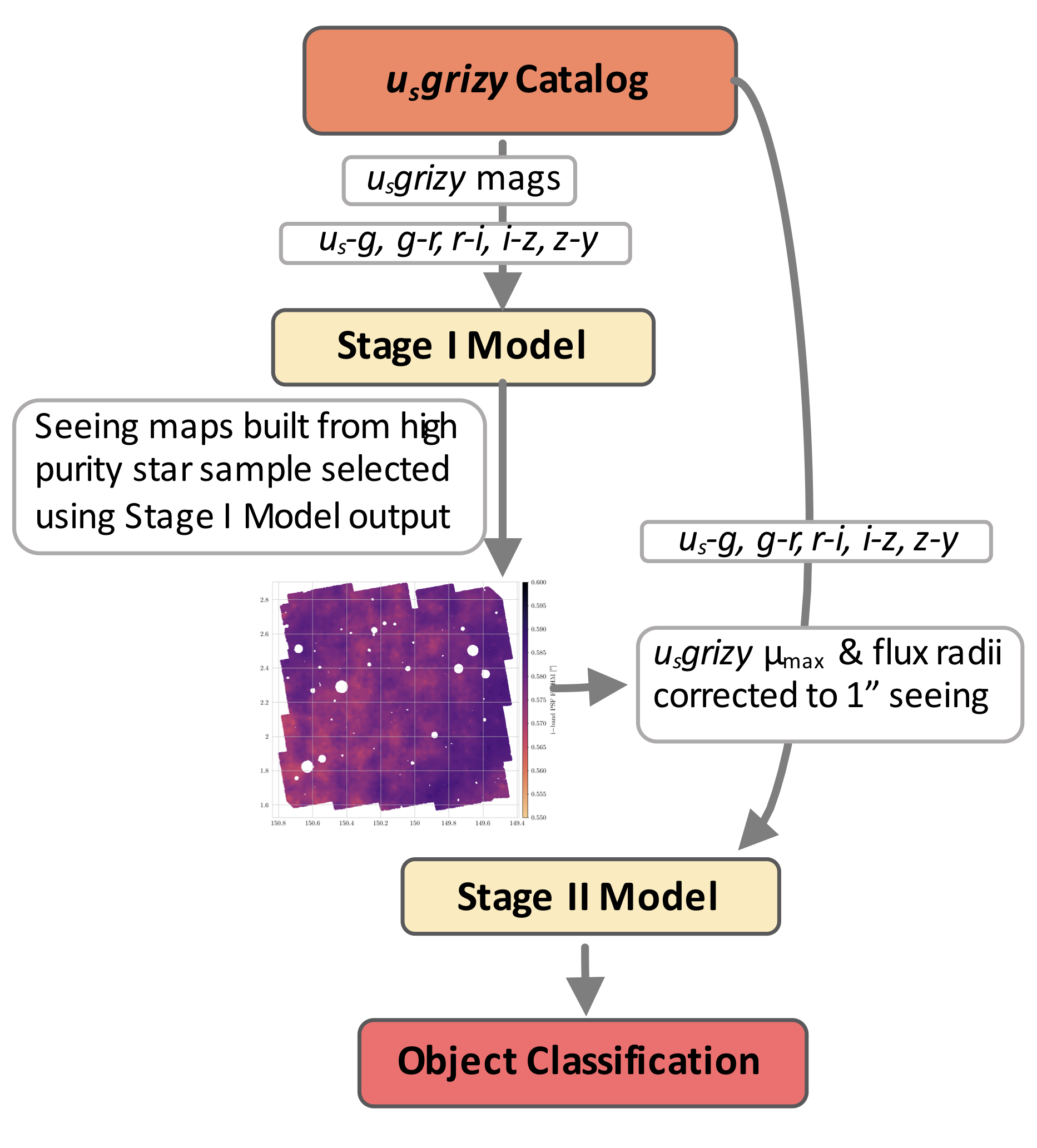}
\par\end{centering}
\caption{Schematic of the two-stage model application.  Model input features are taken from the {\tt SExtractor}-based catalog described in Sec.~\ref{sec:CLAUDSHSCdata}; here $u_s$ is the alternate notation (based on the {\tt u\_S} in the {\tt ASCII} catalog) for $u^*$.
 The Stage I Model, which is described in Sec.~\ref{sec:model1}, is used to generate seeing maps using a high-purity sample of stars selected on the basis of their magnitudes and colors alone. The seeing maps (illustrated with a heatmap in this Figure) are then used to convert measured morphological quantities (maximum surface brightnesses, $\mu_{\mathrm{max}}$, and flux radii) of all objects to a standardized seeing.  These standardized morphological measurements, along with colors, are then used to train the Stage II Model (Sec.~\ref{sec:model2}) that performs the final classification. } \label{fig:fullmodel}
\end{figure}

In this work we describe an object classification pipeline we developed for our CLAUDS+HSC-SSP dataset that takes a catalog of data derived from ground-based imaging as input and outputs the probability that each object belongs to each of the classes under consideration using an ensemble of Gradient Boosted Tree models. The pipeline uses a two-step process to categorize objects. Our stage I model (described in Section \ref{sec:model1}) is trained to select easily identifiable point sources in the catalog. The morphologies of these objects are used to map the point spread function (PSF) over the catalog area. This PSF map is then used to homogenize observations of all objects for use in our stage II model (described in Section \ref{sec:model2}), which outputs the final classification probability for each object. This pipeline is depicted in Figure \ref{fig:fullmodel}.

\section{Data}\label{sec:data}

\subsection{The CLAUDS+HSC-SSP data}\label{sec:CLAUDSHSCdata}

Our goal is to classify objects in the CLAUDS+HSC-SSP  catalog that probes the sky to an unprecedented combination of area and depth.  This catalog is based on images from the Subaru Hyper Suprime-Cam (HSC) $grizy$ Subaru Strategic  Program \citep{hscssp} internal data release s16a (ultradeep deep-depth layer) and the complementary CFHT Large Area U-band Deep Survey \citep[CLAUDS, ][]{clauds}.   These data are described in detail in the aforementioned survey description papers and our catalog creation procedure is in \citet{clauds}, so here we only give a short overview.  

Briefly, the CLAUDS+HSC-SSP catalog covers $\sim$18.60~deg$^2$ in the HSC-SSP Deep and UltraDeep fields.  The data consist of CFHT/MegaCam $u$ and/or $u^*$ and Subaru/Hyper Suprime-Cam $g, r, i, z, y$-band imaging. The $u$ and $u^*$ filters (collectively referred to as the $U$ bands) have slightly different central wavelengths, and thus, where these data overlap spatially, they are kept distinct both at the image stacks and at the catalog level.  The CLAUDS+HSC-SSP data are split into four roughly equal-sized, widely-separated Deep fields on the sky: E-COSMOS, XMM-LSS, ELAIS-N1, and DEEP2-3.  Within the first two of these fields are located the COSMOS and SXDS UltraDeep regions of $\sim$1--1.5~deg$^2$ each. The catalog creation procedure is described in \cite{clauds} and \cite{claudsUVLF}.  Briefly, the individual $U$ images were first transformed to the HSC coordinate system and stacked,  and then object detection and photometry was performed with SExtractor \citep{sextractor} in dual-image mode using a combined $Ugrizy$ (where $U$ means $u$, $u^*$, or both when both are available) $\Sigma$SNR image for object detection \citep[see Section 3.1.2 in][]{clauds}.  Our classification method uses photometric (magnitudes and colors) and simple shape (flux radii and $\mu_{\mathrm{max}}-$mag, where $\mu_{\mathrm{max}}$ is central surface brightness) measurements recorded in our CLAUDS+HSC-SSP catalog. 

We develop and demonstrate our approach to object classification in the CLAUDS+HSC-SSP catalogs using data in the COSMOS and SXDS fields. These two fields are ideal for this purpose as they contain extensive ancillary data -- including  including Hubble Space Telescope (HST) imaging over most of COSMOS and parts of SXDS  -- and which reach a depth of $i_\mathrm{AB}\sim26.5$.   While the CLAUDS dataset includes two $U$ filters ($u$ and $u^*$), this paper illustrates our approach using the $u^*$ data as these are available in both COSMOS and SXDS.  We note that for the purpose of object classification, the distinction between $u$ and $u^*$ is immaterial.  

\subsection{Training and testing data}

HST imaging is essential for training and testing our classification method as spectroscopic star/galaxy classification is not possible at the deep depths reached by the CLAUDS+HSC-SSP dataset.  We match our $u^*grizy$ COSMOS field catalog to that of \citealt{acsCat} (hereafter L+07.) The L+07 catalog includes magnitudes and central surface brightnesses measured in the F814W band in deep images from the HST's Advanced Camera for Surveys. The catalog covers 1.64 deg$^2$ of the COSMOS field and includes $1.2\times 10^6$ objects, complete to F814W=26.5. The high-resolution HST imaging allows unambiguous differentiation of stars and galaxies much fainter than in ground-based images affected by atmospheric seeing. The L+07 catalog includes a star/galaxy classification based on the position of objects in the  $\mu_{\mathrm{max}}$--mag plane that is reliable to F814W=25. Strictly speaking, this classification is between resolved and unresolved morphologies, although we take these to be synonymous with stars and galaxies. We apply the L+07 F814W<25 object star/galaxy labels to their position-matched counterparts in the $u^*grizy$ catalog and use this sample (which includes 620224 galaxies and 20337 stars), split into sub-samples, to train and test our binary star-galaxy classification model (Section \ref{sec:model2}).

\subsection{Data for exploration of star/galaxy/AGN classification}

Finally, to perform a preliminary exploration of our method's ability to classify AGN in addition to stars and galaxies, we also train multi-class models to distinguish stars, galaxies and AGN (Section \ref{sec:stargalagn}). Our AGN sample comes from the xray-selected, multiwavelength catalog of \citet[][hereafter B+10]{brusaCat} that covers 2~deg$^2$ of the COSMOS field. B+10 includes a complete sample of spectroscopically confirmed type I and II AGN down to $i_{AB}=21$. As with the L+07 catalog, we match the B+10 catalog to our $u^*grizy$ sample and apply the AGN labels to matched objects. Our matched sample includes 360 Type I AGN and 310 Type II AGN. We note that the B+10 sample is somewhat noisy as the position matching of x-ray and optical sources is limited by the low spatial resolution of XMM-Newton. Additionally, the Type II AGN sample of B+07 is characterized by a lack of observed narrow line features and L$_{|2-10keV|}>10^{42}$ erg s$^{-1}$. Using this simple threshold it is possible that other types of x-ray sources are misidentified as Type II AGN. Despite these limitations, this AGN sample serves as a useful first exploration of potential future developments of our method.

\section{Machine Learning Background}\label{sec:mathstuff}

\subsection{ Gradient Boosted Trees}\label{sec:stargalml}

Myriad algorithms exist for classifying objects based on tabular data. Here, to build our classification model, we use {\tt XGBoost} \citep{ChenXGB}, an  efficient and powerful open-source implementation of the gradient-boosted tree (GBT) algorithm \citep{friedman2001greedy} that has found application in a variety of problems in astronomy (e.g.\ \citealt{Tamayo2016, Bethapudi2018, Yi2019, Jin2019}). The model's input is the set of features, the observed properties of each object in the catalog (here: colors, magnitudes, central surface brightnesses and several flux radii in all available bands --- see Sec.~\ref{sec:ClassificationModel}). A GBT classification model consists of an ensemble of decision trees, each of which predicts the type of an object based on a series of decisions (represented by nodes in the tree) based on a threshold in one of the input features. A random forest model creates a number of such trees using different subsamples of the training data and input features. Training each tree consists of selecting the optimal thresholds for making the decision at each of its nodes to optimize the final prediction output by the tree. A random forest model produces a single output by having the ensemble of trained trees `vote' \citep{Breiman2001}. The GBT algorithm improves on (\emph{boosts}) the random forest concept by building the trees sequentially and keeping track of the error in the prediction output by the existing ensemble; the tree added at each step is tuned not to independently predict object classes, but to do so while compensating for the mistakes that the ensemble would make without it. 

\begin{table*}
	\centering
	\caption{Optimal {\tt XGBoost} hyperparameter values that we determined by Bayesian Optimization and then used for the four models: the Stage I Model used for PSF homogenization and the three Stage II Model classifiers we produced:  2-Class (star/galaxy), 3-Class (star/galaxy/AGN Type I), and 4-Class (star-galaxy/AGN Type I/AGN Type II)).  For descriptions of the parameters see XGBoost documentation at {\tt https://xgboost.readthedocs.io/en/latest/parameter.html}}
	\label{tab:xgbparams}
	\begin{tabular}{lccccc} 
		\hline
			&	Model I (PSF Homogenization)	&	2-Class	&	3-Class	&	4-Class	\\
			\hline
	Best eval. metric.		&	0.98354	&	0.99946	&	-0.00714	&	-0.0401	\\
{\tt	colsample\_bytree	}	&	0.3649	&	0.869	&	0.8608	&	0.519	\\
{\tt	gamma	}	&	0.9166	&	0.1364	&	0.0417	&	0.2302	\\
{\tt	learning\_rate	}	&	0.289	&	0.2336	&	0.2198	&	0.1876	\\
{\tt	max\_delta\_step	}	&	5.8564	&	1.6102	&	7.725	&	2.7982	\\
{\tt	max\_depth	}	&	6	&	8	&	7	&	6	\\
{\tt	min\_child\_weight	}	&	0.6426	&	1.3555	&	1.509	&	0.6967	\\
{\tt	n\_estimators	}	&	165	&	180	&	135	&	127	\\
{\tt	reg\_alpha	}	&	0.3282	&	1.35	&	0.9883	&	1.7212	\\
{\tt	reg\_lambda	}	&	1.7753	&	1.6047	&	0.5909	&	1.2923	\\
{\tt	subsample	}	&	0.8121	&	0.7646	&	0.573	&	0.7142	\\
{\tt	objective	}	&	binary:logistic	&	binary:logistic	&	multi:softprob	&	multi:softprob	\\
{\tt	eval\_metric	}	&	auc	&	auc	&	mae	&	mae	\\
		\hline
	\end{tabular}
\end{table*}

\subsection{SMOTE: Synthetic Minority Over-sampling TEchnique}

Any flux-limited, deep, photometric sample will contain vastly more galaxies than stars and vastly more stars than AGN. In other words, the classes we wish to consider are highly imbalanced. Class imbalance in training data can affect a model's sensitivity to minority classes because the overall penalty for misclassifying them is small.

Many methods exist to rebalance classes in data. Cost functions which can weight classes differently can be used, the majority class can be undersampled by removing objects from the training sample, or the minority class can be oversampled by duplicating objects in the training sample, or generating new objects. In training our {\tt XGBoost} model, we use the Synthetic Minority Over-sampling Technique (SMOTE, \citealt{SMOTE}) to generate synthetic examples of stars and AGN to match the number of galaxies in the training data. The SMOTE algorithm works in feature space, finding the nearest neighbour of each minority class object. It then generates a synthetic example of the class by randomly selecting a value for each feature on the line segment connecting the object and its nearest neighbour. The synthetic object is then added to the dataset and treated no differently from the {\emph{real}} examples during training.  For clarity, we note here that only data used in model training were balanced using the SMOTE algorithm;  those used for testing the performance of our classifiers were not.

\subsection{Hyperparameter Selection with Bayesian Optimization}

Most machine learning techniques have a set of parameters which specify the overall configuration of the model to be trained. Since these parameters are fixed before the specific model is trained, they are commonly called hyperparameters to distinguish them from the lower-level model parameters tuned during training. {\tt XGBoost} includes various hyperparameters that can be adjusted to improve the performance of a trained model. These can be divided in three main classes: general, tree booster and learning hyperparameters. The general hyperparameters adjust properties such as the number of computer cores to be used during training as well as the level of output expected from the code. Tree booster hyperparameters handle the complexity of the decision trees as well as how they are built. These parameters also include {\tt XGBoost}'s selection of regularization terms that prevent the model from overfitting to noise in the data. Finally, the learning hyperparameters specify ways to evaluate training performance. To do this a loss function, also known as objective function, is specified as well as an evaluation metric to assess the error between the predicted and real values of the training data during each step of the process. Various methods to find the best combination of hyperparameters exist but for this work we use Bayesian Optimization \citep{bayesOpt} to tune the parameters of each {\tt XGBoost} classifier. This method models the performance of the classifier with different combinations of hyperparameters as samples from a Gaussian process. This framework allows the algorithm to determine which hyperparameters have the greatest influence on performance so it can select new hyperparameter combinations to test which are likely to significantly improve overall performance. This strategic search allows the algorithm to converge on an optimal parameter set more quickly than a grid search. Table \ref{tab:xgbparams} lists the optimal {\tt XGBoost} hyperparameters selected for the four models we produced (the Stage I Model used for PSF homogenization and three Stage II Models:  the star/galaxy, star/galaxy/AGN Type I, and star/galaxy/AGN Type I/AGN Type II classifiers).

\subsection{{\tt XGBoost}  Ensembles}\label{sec:ensembleTraining}

{\tt XGBoost} improves on the random forest concept through an iterative process (Sec.~\ref{sec:stargalml}). In each model training iteration we divide our catalog into training and test samples which are stratified by class; the two samples contain an equal number of true examples of each class. These subsamples of the data are commonly referred to as 'folds'.

\begin{figure}
\begin{centering}
\includegraphics[width=0.48\textwidth]{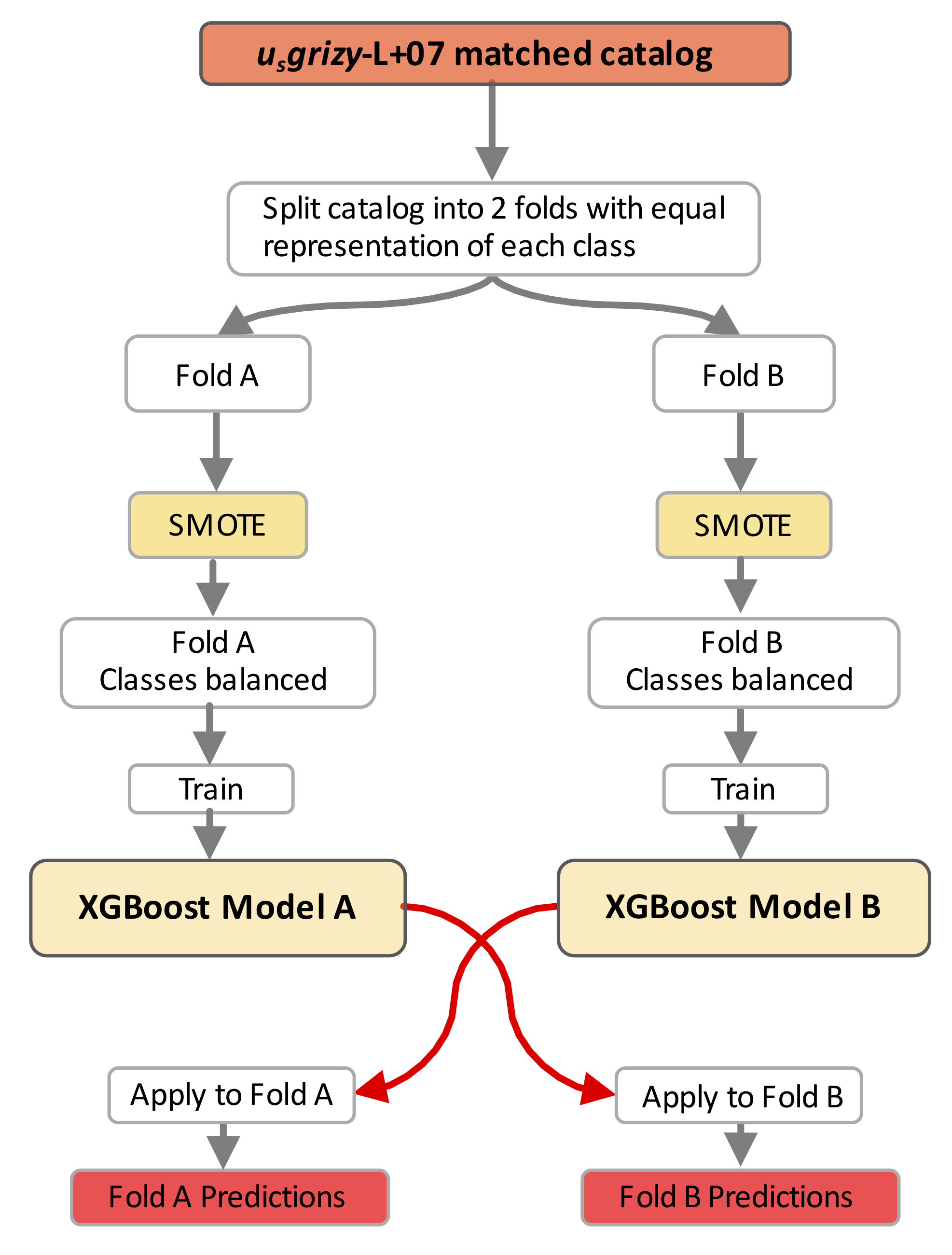}
\par\end{centering}
\caption{Flowchart depicting the procedure used to train each of the {\tt XGBoost}  models.  The \citet{clauds} CLAUDS+HSC catalog is matched with star/galaxy labels from L+07 and serves as input that is split into two independent folds.  Within each fold, the data are SMOTEd against class imbalance, the {\tt XGBoost} model is trained on that fold, and then applied to the other fold to make classification predictions.  Note that performance testing is  done subsequent to the steps shown in this diagram using data that were not SMOTEd.}
\label{fig:modeltraining}
\end{figure}

To enhance the performance of our classifier we train many {\tt XGBoost} models using different stratified splits of the catalog to build an ensemble.  We combine the set of predictions output for each object by the ensemble members to determine our final classification for each object. When applying the ensemble model to other datasets, the full model ensemble is used. However in evaluating the performance of the model ensemble, we calculate the final output for each object using only those models which were trained on samples which did not include the given object to prevent information leakage. Our model training procedure is depicted in Figure \ref{fig:modeltraining}.

\subsection{Evaluation Metrics}\label{sec:EvaluationMetrics}
The output of the {\tt XGBoost}  model is a set of probabilities representing the likelihood that the object under consideration belongs to each possible class. We use several metrics to evaluate the quality of the predictions output by the model for the binary star/galaxy classification task (c.f. \citealt{Sevilla-Noarbe2018}). Our performance metrics can be divided into two classes: those which evaluate the continuous values output by the models (e.g. the probability that a given object is a star), and those that require that the probabilities be mapped to discrete classes before evaluation (e.g. purity and completeness). The class probabilities are converted to labels by applying a threshold probability: objects with probabilities higher than the threshold are labelled as stars, those with probabilities below the threshold are labelled as galaxies.  Once the probabilities have been converted to labels, we discuss model performance in terms of the numbers of true positives (TP, objects correctly predicted as belonging to a class), true negatives (TN, objects correctly predicted as not belonging to the class), false positives (FP, objects incorrectly predicted as belonging to the class), and false negatives (FN, objects incorrectly predicted as not belonging to the class).

\subsubsection*{Mean-Squared Error}
The Mean-Squared Error (MSE) considers the square of the difference between the output probability and true class label for all objects in the test sample \citep{mse}:

\begin{equation}
\text{MSE}(y, \hat{y}) = \frac{1}{n_\text{objects}} \sum_{i=1}^{n_\text{objects}} (y_i - \hat{y}_i)^2,
\end{equation}
where $y_i$ and $\hat{y}_i$ represent the true class and the probability of belonging to that class that is output by the model for the i$^{\mathrm{th}}$ object.

\subsubsection*{Calibration Error}
The Calibration error (CAL) measures the quality of the probabilities output by the model. If the model is well-calibrated, the output probability that a given object belongs to a class should be equal to the proportion of objects belonging to the class for which the model outputs that probability. To quantify this relationship, the output probabilities are often binned so that the average probability for objects in a bin can be compared to the ratio of classes represented in that bin \citep{calibrationError}. We adopt the method used by \citet{sgspConvNN2017}, calculating the difference between the predicted probability and class frequency in overlapping probability bins and taking the mean difference (weighted by the number of objects in each bin) as the model's CAL.

\subsubsection*{Logistic Loss}
The logistic loss (also known as log loss or cross-entropy) is the negative log-likelihood of the true class given the output of the classifier. It is defined as

\begin{equation}
-\frac{1}{N}\sum_{i=1}^{N}\sum_{j=1}^{M}y_{ij}\textrm{log}\,p_{ij},
\end{equation}
where $N$ is the number of samples, $M$ is the number of classes, $y$ is a binary term that indicates if class $j$ is correctly assigned, and $p$ is the probability of sample $i$ being of class $j$. 

The logistic loss is often used as an evaluation metric because it penalizes incorrect classifications heavily.

\subsubsection*{AUC}
The receiver operating characteristic (ROC) curve is a plot of the true positive rate as a function of the false positive rate as the threshold used to map the probability to discrete labels is varied. The area under the ROC curve (AUC) is a commonly used measure of the overall performance of a binary classifier that is insensitive to class imbalance.

\subsubsection*{Purity \& Completeness}
The purity of the set of objects predicted to belong to a given class  (also known as the precision or positive predictive value of the model) is defined as TP/(TP+FP). The completeness of the sample of objects predicted to belong to the class (also known as the recall or sensitivity of the model) is defined as TP/(TP+FN).

\subsubsection*{F1 Score}
The F1 score is the harmonic mean of the purity and completeness of objects predicted to belong to a class. It is calculated for a given class, as
\begin{equation}
\mathrm{F1}=\frac{2}{1/{\mathrm{Completeness}}+1/{\mathrm{Purity}}}=\frac{\mathrm{TP}}{2\times \mathrm{TP+FP+FN}}.
\end{equation}
We compute an unweighted and a weighted F1 score for the binary classifier. In both cases, the F1 score for both classes considered by the model (stars and galaxies) are computed separately and averaged to produce the final score. The unweighted F1 score is the simple mean of the two scores and the weighted F1 score is the average weighted by the number of objects in the test data belonging to each class.


\section{Classification Model}\label{sec:ClassificationModel}

\subsection{Input Features}\label{sec:features}

\begin{figure}
\begin{centering}
\includegraphics[width=0.48\textwidth]{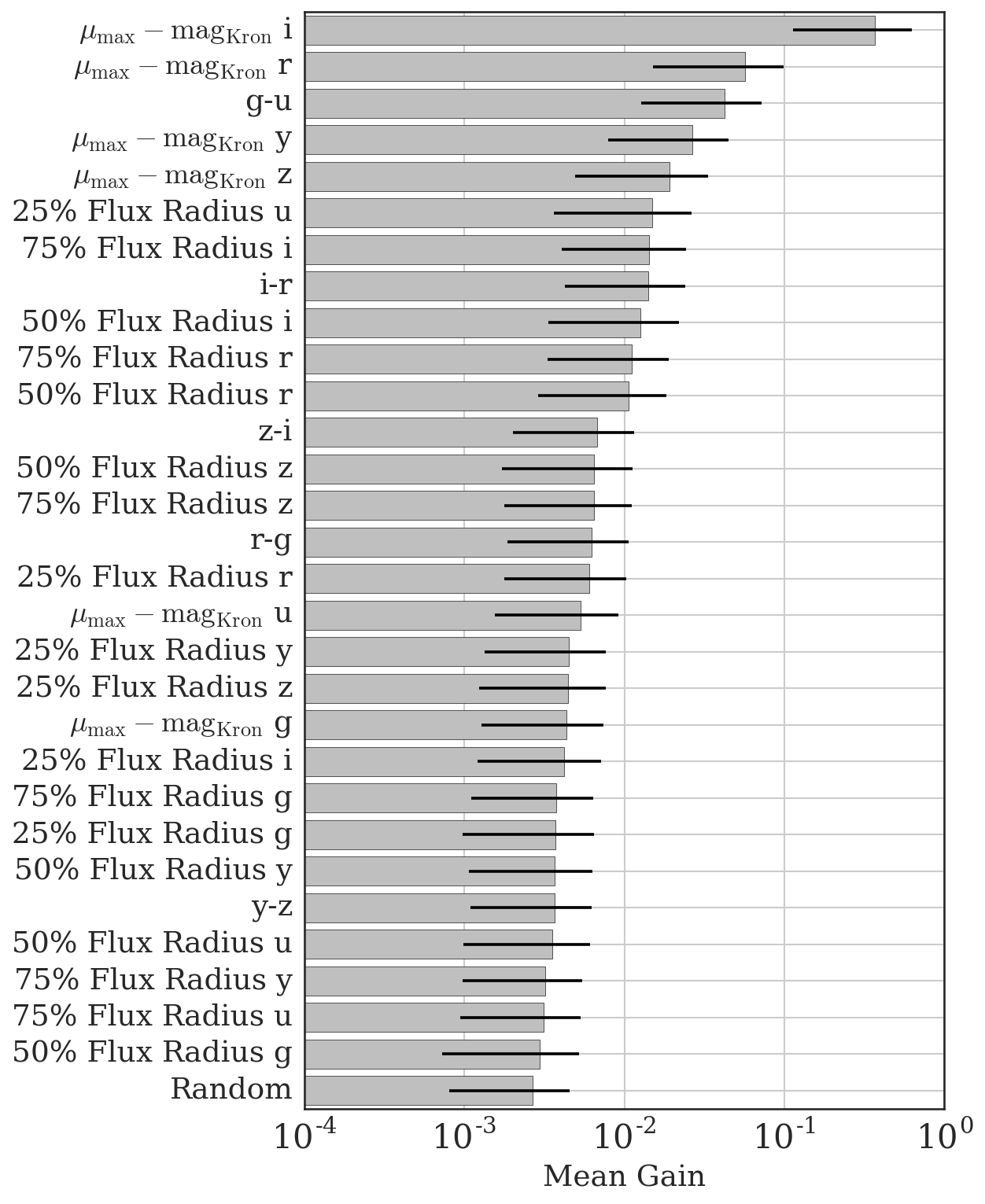}
\par\end{centering}
\caption{Mean information gain and standard deviation associated with each input feature after 100 {\tt XGBoost}  models are trained on different stratified splits of the sample. The input features are explained in Sec.~\ref{sec:features} and are presented in this Figure in decreasing order of main information gain.  $\mu_\mathrm{max}-\mathrm{mag\_{Kron}}$ and flux radius features are corrected to 1" seeing using the stage I model, as described in Section \ref{sec:model1}. All of the features used by the model are more informative than random noise (last row).}
\label{fig:featureImportance}
\end{figure}

As discussed in Section \ref{sec:intro}, stars and galaxies can be separated by their physical extent in an image and by the shapes of their SEDs.  This can be done using measurements produced by the {\tt SExtractor} \citep{sextractor} software (e.g., \citealt{Soumagnac2015}).
With this in mind, we select a number of features from our CLAUDS+HSC-SSP {\tt SExtractor} catalog likely to be useful in distinguishing the two samples: Kron magnitude ({$\mathrm{mag_{Kron}}$}: {\tt SExtractor} parameter {\tt MAG\_AUTO}), the difference between maximum surface brightness and Kron magnitude, ($\mu_\mathrm{max} - \mathrm{mag_{Kron}}$: {\tt SExtractor MU\_MAX-MAG\_AUTO}), and the radius enclosing 25\%, 50\%, and 75\% of total flux ({\tt FLUX\_RADIUS} with  {\tt PHOT\_FLUXFRAC} set to 0.25, 0.5, 0.75) measured in each of the $u^*grizy$ bands, as well as photometric colours $u^*-g$, $g-r$, $r-i$, $r-z$, $z-y$ from {\tt FLUX\_ISO} measurements. The full set of features is listed in Figure \ref{fig:featureImportance}. We consider the Kron magnitudes and colours to be photometric features, and maximum surface brightnesses and flux radii to be morphological features. 

We note significant variation in the calibration of $\mu_\mathrm{max}$ and {\tt FLUX\_RADIUS} measurements caused by seeing differences in different areas of the survey. If not accounted for, these differences would make a model trained on part of the catalog useless when applied to the remainder. Section \ref{sec:model1} describes our procedure for homogenizing the morphological features to a standard seeing of 1\arcsec, while the rest of this Section (Sec.~\ref{sec:model2}--\ref{sec:stargalxmm}) uses morphological features which have undergone this standardization.\\

We correct flux measurements for Galactic extinction using the prescription of \cite{schlegel_maps_1998}. This correction is critical for measuring accurate galaxy fluxes, but applying it to all catalog objects may produce biased fluxes in stars which are not observed through the full extent of the Galaxy's dust screen. The vast majority of these stars will, however, appear very bright, and their classification should be driven by morphological features.

We confirm that all of the selected features are informative to the classification by adding a random noise column to the input features (Fig. \ref{fig:featureImportance}). We then repeatedly train the {\tt XGBoost} model and track the average information gain that results when each feature is used by a decision tree to split the sample. The information gain compares the entropy of the sample, S (where entropy is broadly defined as the uncertainty in the possible outcomes), to be split based on the feature to the average entropy of the two child groups created once the split is made \citep{entropy}. The entropy of a group is defined as,
\begin{equation}
E=\sum_{i=1}^{n_\text{classes}} -p_i log_2p_i,
\end{equation}
where $p_i$ is the probability that an object in the group belongs to the $i^{th}$ class. The information gain, G, when a split is made is then
\begin{equation}
G=E(S_\mathrm{parent})-0.5\times (E(S_\mathrm{child,1})+E(S_\mathrm{child,2})),
\end{equation}
representing the drop in entropy. This provides a measure of each feature's usefulness in discriminating between the classes.

 Figure \ref{fig:featureImportance} shows the mean and standard deviation of the information gain associated with each feature used in the final model after training 100 {\tt XGBoost} models. As expected, $\mu_\mathrm{max}-\mathrm{mag_{Kron}}i$ is the most useful feature:  when seeing conditions are best, $i$-band observations are carried out by the HSC-SSP observing team. However, all of the other features 
 contribute higher drops in entropy than the random noise feature, justifying their inclusion in the set of input features.

\subsection{Stage I: PSF Homogenization}\label{sec:model1}

We expect measurements that characterize morphology to be among the most useful features for classifying stars and galaxies. However, seeing conditions can vary significantly over a ground-based survey.  Although in the CLAUDS+HSC-SSP fields seeing is very good and always better than 1\arcsec (ranging from a median $\sim$0.9\arcsec\ in the $U$ bands to $\sim$0.6\arcsec\ in $i$-band; \citealt{clauds, hscsspPDR1}), it varies between bands and from location to location.  Consequently, a model trained on data from one field will not be directly applicable to the full survey area without additional (pre)-treatment. 

Once a sample of stars has been identified, their half-light radii can be used to map the seeing over the catalog area if the functional form of the point spread function (PSF) is known. The star sample used to compute this map need not be as deep or complete as the full catalog. It is only important that the sample has high purity and a distribution on the sky that is dense enough to trace the variation of the PSF. 

To build seeing maps in each band, we train our first model (Stage I Model) using only features which are insensitive to different seeing conditions ($u^*-g$, $g-r$, $r-i$, $i-z$, $z-y$ colours computed in 2" apertures, and $u^*grizy$ Kron magnitudes) by applying the training method described in Section \ref{sec:ensembleTraining} to the L+07-matched $u^*grizy$ catalog. This gives us a first estimate of the probability that each object is a star. Using the sample of known stars, we determine the threshold for this probability that will produce a sample of stars with 99\% purity at magnitudes $19 < i_{AB} < 21.5$. The half-light radii ($r_{50\%}$) of this sample of probable stars are used to infer the width of the PSF at their positions on the sky assuming a Gaussian PSF (FWHM$=2\times r_{50\%}).$ We use these measurements to interpolate the PSF width at the position of each object in the catalog (in each band) by taking the median FWHM of all probable stars within a radius of 0.05 deg from its center. The PSF values for each object are then used to correct the object's morphological features to an effective seeing of 1". With our assumption of a Gaussian PSF, $\mu_\mathrm{max}$ is scaled by (FWHM/1")$^2$ and {\tt FLUX\_RADIUS} measurements are scaled by (FWHM/1") to preserve total fluxes.

\subsection{Stage II: Binary Star/Galaxy Classification}\label{sec:model2}

After applying the PSF homogenization model (the Stage I Model) to the $u^*grizy$ data, we train our final {\tt XGBoost}  model ensemble using the full set of input features listed in Figure \ref{fig:featureImportance}. The same procedure used to train Model I (depicted in Figure \ref{fig:modeltraining}) is used to generate 100 {\tt XGBoost}  models and 50 blind predictions for each object in the $u^*grizy$-L+07 matched catalog which give the probability that each object is a star.

We use the metrics described in Section \ref{sec:mathstuff} to evaluate the performance of the model and list their values in the second column of Table \ref{tab:othersamplestable2}. The model achieves a high AUC score of 0.9974. While {\tt SExtractor}'s {\tt CLASS\_STAR} classifier has in recent years been surpassed by other methods (e.g. \citealt{Sevilla-Noarbe2018}), it remains a widely-used gold standard. In the $i$-band (where seeing is best) and for the same sample {\tt CLASS\_STAR}  gives an AUC score of 0.9043; so the {\tt XGBoost} ensemble provides a significant improvement. The low calibration error, CAL, of 0.0163 indicates that the probabilities output by the model are truly representative of the probability that individual objects are stars.

\begin{figure}
\begin{centering}
\includegraphics[width=0.47\textwidth]{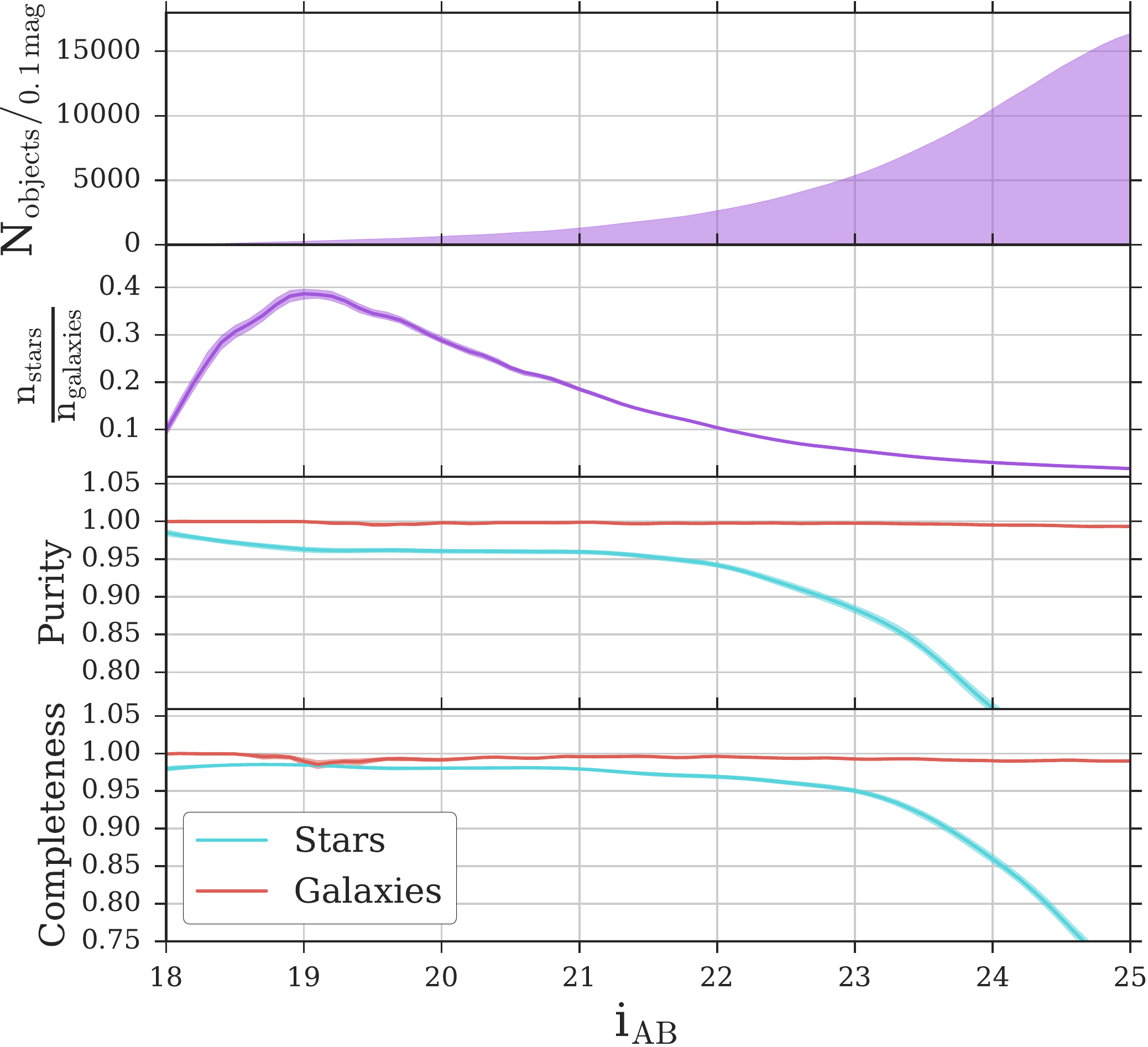}
\par\end{centering}
\caption{ Purity and completeness are calculated at the threshold that maximizes the F1 score (threshold=0.879). Shaded regions show bootstrapped 68\% confidence intervals. This result is specific to the COSMOS field catalog, results are likely to vary depending on Galactic latitude. \label{fig:performanceInCosmos}}
\end{figure}

We report the F1 scores, purity and completeness at the threshold that maximizes the weighted F1 score for the sample (0.879). The unweighted F1 score of 0.9158 is  lower than the weighted F1 score because of the significant class imbalance between stars and galaxies. The weighted F1 score (0.9981) is sensitive to the rate at which the classes occur in the test sample and de-emphasizes the performance of the model on stars when determining the optimal threshold for converting the probability output by the model to a discrete classification. This leads to very high completeness and purity in the predicted sample of galaxies ($\mathrm{P_{galaxies}}=0.9974$, $\mathrm{C_{galaxies}}=0.9982$) and slightly lower values for stars ($\mathrm{P_{stars}}=0.9635$, $\mathrm{C_{galaxies}}=0.8727$.) The method used to select this optimal threshold can be varied to select a sample that is optimized -- in terms of purity or completeness -- differently for different science cases.  Figure \ref{fig:performanceInCosmos} shows how the purity and completeness of the star and galaxy samples change as a function of $i_\mathrm{AB}$ at the threshold noted above. Both quantities drop very slightly for galaxies toward $i_\mathrm{AB}\sim 25$, while they drop more significantly  for stars at faint magnitudes where vastly more galaxies than stars are observed.

\subsection{Generalization to fainter objects}\label{sec:stargalgen}
The performance of the {\tt XGBoost} model described in Section \ref{sec:model2} is only applicable to data similar to those used in training. For very deep surveys such as CLAUDS+HSC-SSP, it's important that we understand how well the model classifies objects fainter than those appearing in the training data.
To explore this issue, we consider various values of a threshold brightness $i_{\mathrm{AB,max}}$, training an {\tt XGBoost}  model on the objects with $i_{\mathrm{AB}}<i_{\mathrm{AB,max}}$ and then considering its performance on the sample of objects with $i_{\mathrm{AB,max}}$ fainter than $i_{\mathrm{AB,max}}$. Here, we fix the {\tt XGBoost} hyperparameters to the optimal values found for the sample as a whole, and determine the optimal threshold for converting the model prediction to a binary classification by maximizing the weighted F1 score for the $i_{\mathrm{AB}}<i_{\mathrm{AB,max}}$ training data.

\begin{figure*}
\begin{centering}
\includegraphics[width=0.5\textwidth]{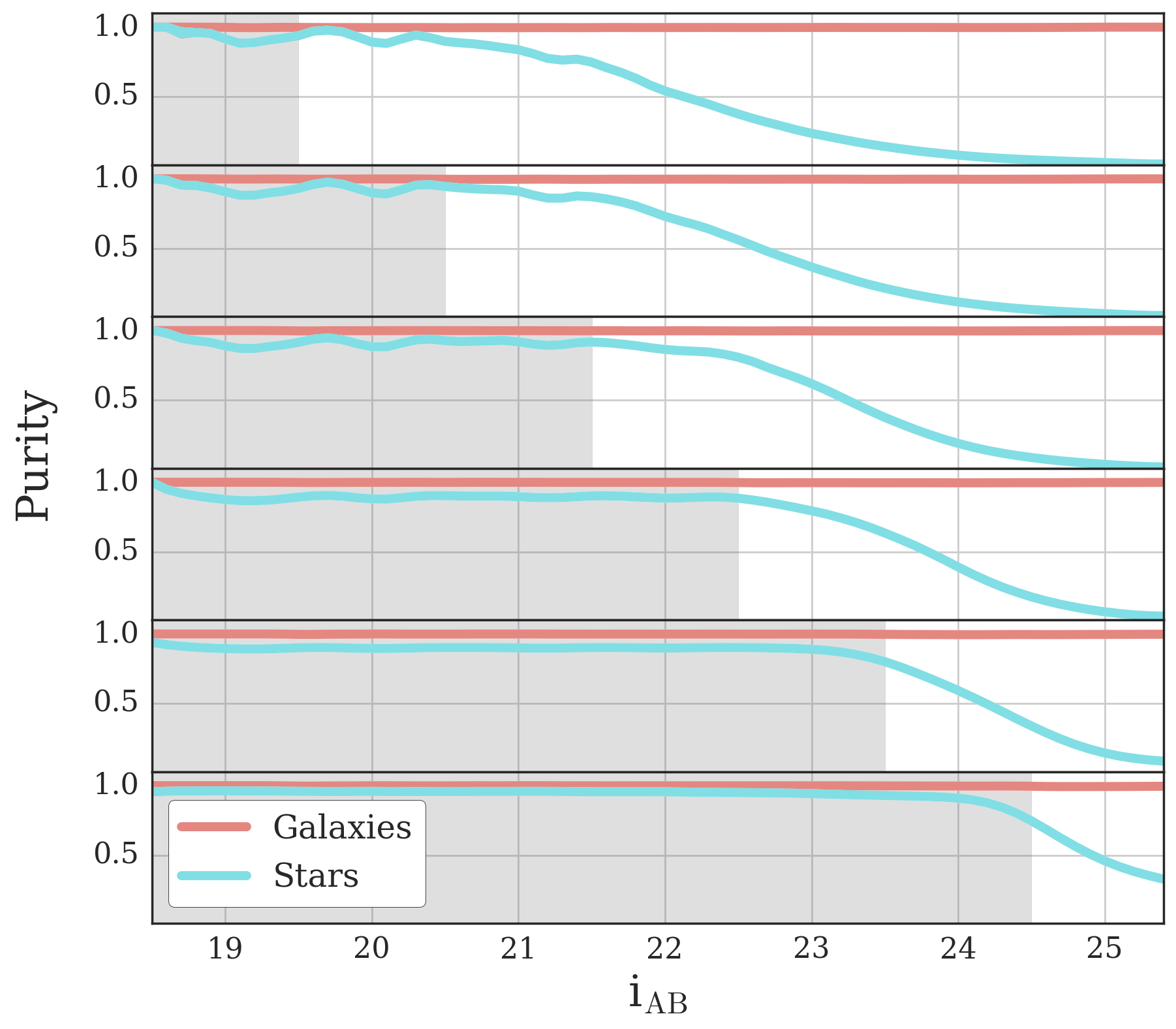}\includegraphics[width=0.5\textwidth]{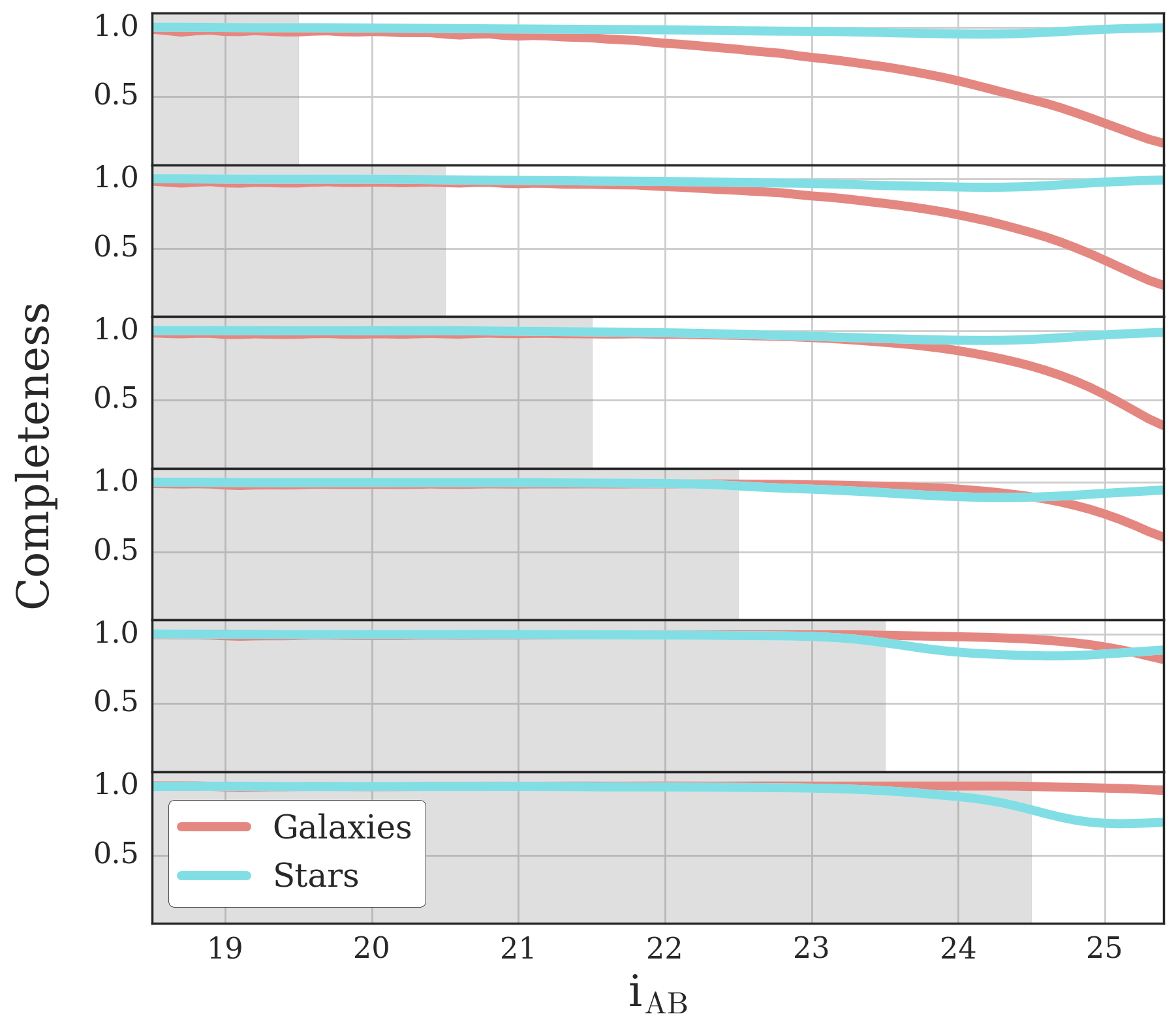}
\par\end{centering}
\caption{Purity and completeness of star/galaxy classification for objects fainter than the training set. In each row of both panels, the shaded region shows the range of magnitudes included in the training sample and the purity of stars and galaxies selected by each model as a function of $i_{\mathrm {AB}}$ magnitude.  \label{fig:purecomp}}
\end{figure*}

\begin{figure}
\begin{centering}
\includegraphics[width=0.52\textwidth]{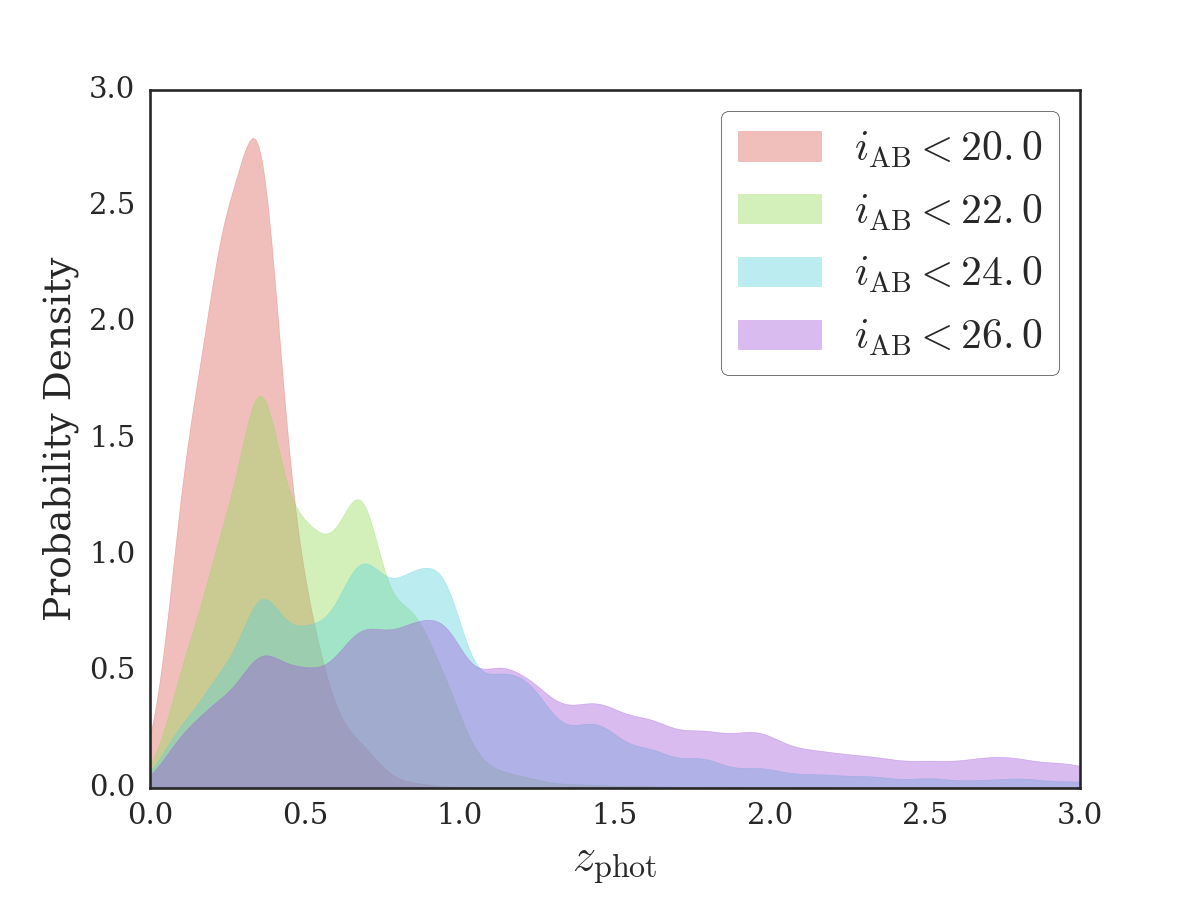}
\par\end{centering}
\caption{Gaussian kernel density estimates of photometric redshifts for galaxies brighter than the $i_\mathrm{AB}$ threshold. Bright samples include few galaxies at high redshift. Classification models trained on bright samples will perform poorly when applied to faint samples composed largely of objects at high redshift.\label{fig:zHistI}}
\end{figure}

Figure \ref{fig:purecomp} shows the purity and completeness of the star and galaxy samples as a function of $i_\mathrm{AB}$ for models trained on samples with progressively fainter $i_{\mathrm{AB,max}}$. The completeness of stars is effectively independent of the depth of the training set. This makes sense in that the identification of stars largely relies on the morphological features ($\mu_{\mathrm{max}}-i_\mathrm{Kron}$, and {\tt FLUX\_RADIUS}) and the colours of nearby, bright stars are the same as those of more distant, faint stars of the same stellar type. Meanwhile, faint galaxies are small on the sky and have morphologies very similar to those of stars, so to distinguish them from stars, the model must rely primarily on their colour features. The observed colours of galaxies depend on their redshifts. To investigate how the accuracy of the classifier changes with redshift, we position-match our $u^*grizy$ catalog with the multi-wavelength COSMOS2015 photometric redshift catalog of \cite{laigle2017}. Figure \ref{fig:zHistI} shows that at bright magnitudes, high redshift galaxies are very rare in our sample. If the training sample of the {\tt XGBoost} model is much shallower than the test sample under consideration, the faint, high redshift galaxies will have no colour-counterparts in the bright sample; the model will then be likely to classify them as stars. This is seen in the dropping completeness of galaxies in the right panel of Figure \ref{fig:purecomp}. The drop in purity of the stars with increasing magnitude seen in the left panel of Figure \ref{fig:purecomp} is driven by the same effect: the sample is contaminated by galaxies whose redshifts (and thus colours) are not present in the training data. 

We conclude that this model can reasonably be applied to objects $\sim1-2$ magnitudes fainter than the training data without a significant drop in completeness for either class. If a \emph{pure} sample of galaxies is required, the model can be relied upon for objects at any magnitude, independent of the depth of the training data. However, if a high-purity sample of stars is desired, extrapolation to fainter magnitudes should be avoided.  For the CLAUDS+HSC-SSP dataset, this means that classifying to $i\sim26-27$AB should generally give very good results. 

More broadly, these results suggest that the most important factor in building an extrapolable model for star/galaxy classification is ensuring that the range of redshifts (observed colours) of the galaxies in the training data are representative of the fainter test sample. When such data are unavailable, it may be advisable to augment the training sample with synthetic galaxy observations to augment the resulting model's robustness.  Exploration of such approaches is left for the future.

\subsection{Model Performance on Different Samples from CLAUDS+HSC-SSP}
\label{sec:stargalxmm}
Depending on the line of sight observed in the Galaxy, we might expect the underlying population of stars to differ significantly from that which the model is trained to identify. To examine this effect, we apply the full, 2-stage model (trained on the $u^*grizy$-L+07 catalog in the COSMOS field) to our catalog in the SXDS field.

\begin{figure}
\begin{centering}
\includegraphics[width=0.48\textwidth]{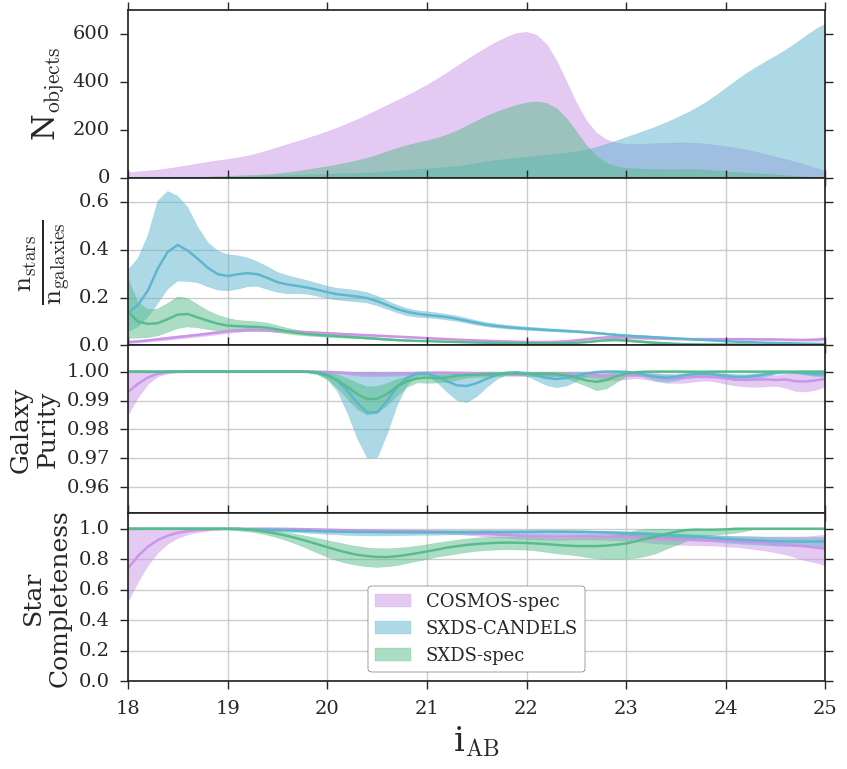}
\par\end{centering}
\caption{Performance of the two-stage model on a 0.07 deg$^2$ patch in SXDS with stars selected morphologically, and spectroscopically selected samples in SXDS and COSMOS. Shaded regions in the bottom three panels show bootstrapped 68 \% confidence intervals. Purity and completeness are calculated using the threshold that maximized the weighted F1 score in the COSMOS $Ugrizy$-L+07 sample.\label{fig:otherData}}
\end{figure}

To evaluate performance in SXDS, we select a sample of \emph{true} stars in two ways. We use $\sim 0.06$ deg$^2$ of HST-ACS data from CANDELS \citep{candels2011} to select stars using the same $\mu_\mathrm{max}-\mathrm{mag}$ technique used by L+07. This sample is complete to $i_\mathrm{AB}\sim 26$ and includes 7859 galaxies and 316 stars. We also use a spectroscopic catalog which includes high-quality observations of 20007 objects in the COSMOS and SXDS fields from numerous spectroscopic surveys (\citealt{zcosmosLilly2007}, \citealt{vipersScodeggio2017}, \citealt{vudsTasca2017}, \citealt{fmoscosmosSilverman2015}, \citealt{mosdefKriek2015}, \citealt{Comparat2015}, \citealt{udszMcLure2013}, \citealt{udszBradshaw2013}).  
The top two panels of Figure \ref{fig:otherData} show the composition of the SXDS CANDELS-selected sample, and of the spectroscopic samples in SXDS and COSMOS. The selection criteria of the spectroscopic samples (designed to measure galaxy redshifts) in both fields mean that they are much brighter and contain many fewer stars than the flux-limited samples.

The bottom two panels of Figure \ref{fig:otherData} show the galaxy purity and star completeness of the samples output by the model (with threshold=0.879) for each dataset as a function of $i_\mathrm{AB}$. The galaxy purity of all samples is very high, indicating that the galaxy sample used in training was representative of the galaxy populations that occur in each of the other datasets. The completeness of stars is similarly high for the SXDS-CANDELS sample where stars were identified using $\mu_\mathrm{max}-\mathrm{mag_{Kron}}$, as they were in the COSMOS field training sample from L+07 and for the spectroscopically-selected sample in COSMOS. In the SXDS spectroscopic sample, the completeness of stars is slightly lower and fluctuates with $i_\mathrm{AB}$. We attribute this difference in performance to the small size of the sample, and over-representation of very bright stars (compared to the COSMOS field training data) therein.

The last 3 columns of Table \ref{tab:othersamplestable2} compare the model's performance on these  datasets with its performance on the $u^*grizy$-L+07 sample. For those metrics that require the selection of a threshold (F1, purity and completeness) we use the value that maximizes the F1 score in the $u^*grizy$-L+07 COSMOS sample on which the model was trained. Whenever a model is applied to a dataset that is in some way different from that on which it was trained, some drop in performance is to be expected. Here, we note small decreases in performance for most metrics considered. Overall, however, the model trained on the COSMOS sample performs well even when applied to another, widely-separated field in the survey.

\begin{table*}
	\centering
	\caption{Comparison of performance metrics of the binary classification model applied to other samples. The F1 score is calculated at the threshold that optimizes the weighted F1 score for the $ugrizy$-L+07 sample. The last four columns refer to the two fields of the CLAUDS+HSC-SSP survey (COSMOS and SXDS) and the imaging  (L+07, CANDELS) or spectroscopic validation samples used.  The performance metrics named in the first column are defined in Sec.~\ref{sec:EvaluationMetrics}.}
	\label{tab:othersamplestable2}
\begin{tabular}{|l|c|c|c|c|}
\hline 
 & COSMOS-L+07 & COSMOS-spec & SXDS-CANDELS & SXDS-spec\tabularnewline
\hline 
AUC & 0.9974 & 0.9859 & 0.9900 & 0.9530\tabularnewline
MSE & 0.0091 & 0.0164 & 0.0367 & 0.0083\tabularnewline
log loss & 0.0368 & 0.0781 & 0.1614 & 0.0388\tabularnewline
CAL & 0.0163 & 0.0192 & 0.0580 & 0.0089\tabularnewline
F1$_{\mathrm{{weighted}}}$ & 0.9918 & 0.9888 & 0.9817 & 0.9948\tabularnewline
F1 & 0.9158 & 0.8205 & 0.7324 & 0.8394\tabularnewline
P$_\mathrm{{galaxies}}$ & 0.9974 & 0.9983 & 0.9977 & 0.9974\tabularnewline
C$_\mathrm{{galaxies}}$ & 0.9982 & 0.9894 & 0.9813 & 0.9973\tabularnewline
P$_\mathrm{{stars}}$ & 0.9635 & 0.7253 & 0.6056 & 0.8351\tabularnewline
C$_\mathrm{{stars}}$ & 0.8727 & 0.9444 & 0.9262 & 0.8438\tabularnewline
\hline 
\end{tabular}
\end{table*}

\section{Star-Galaxy-AGN Classification}
\label{sec:stargalagn}

For many science applications it is also desirable to identify AGN. In this Section we explore the possibility extending our classification procedure to classifying AGN in addition to stars and galaxies. 

Matching our COSMOS field catalog to that of B+10 yielded a sample of spectroscopically confirmed AGN including 360 Type I and 310 Type II. The B+10 is complete to $i_\mathrm{AB}=21$ (all objects in the $u^*grizy$ catalog that had a matched x-ray source were excluded from the binary star/galaxy classification training and test data described in Section \ref{sec:model2}). This AGN sample is relatively small and limited to bright objects; this multiclass modelling exercise should be considered an exploration of what might be possible with improved training data.

We train a 4-class model to distinguish stars, galaxies, Type I and Type II AGN. Using all catalog objects with $i_{AB}<21$, we first apply the PSF homogenization model described in section \ref{sec:model1} and then work with the same input features used in the binary star/galaxy classification model. As in the two-class case, we divide the sample into two stratified folds and use SMOTE to create synthetic examples of the minority classes to correct the class imbalance. We then train two {\tt XGBoost} models with the optimal hyperparameters listed in Table \ref{tab:xgbparams} determined by Bayesian optimization. In the multiclass case, we use a softmax objective function (also known as a normalized exponential function) and the mean error as the evaluation metric. After training, we use the two models to predict the class of those objects in the half of the data that they  were not exposed to during training. 

We repeat this procedure 50 times, to generate a total of 100 models and 50 sets of predicted probabilities for each of the objects used in training. The output of the multiclass classifier for each object is a probability distribution over the set of  possible classes. For each object, we sum the 50 probabilities corresponding to each class and select the class with the highest summed probability as the output class of our ensemble model. 

\begin{figure}
\begin{flushright}
\includegraphics[width=0.4\textwidth]{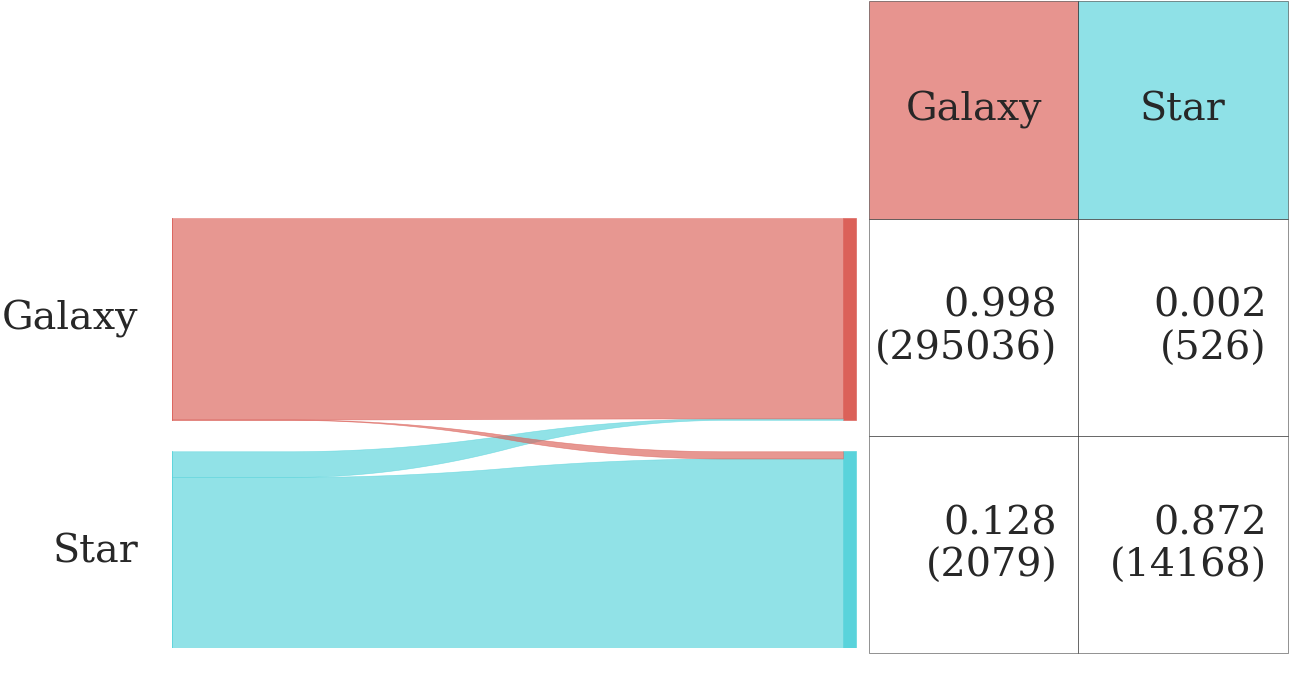}
\includegraphics[width=0.46\textwidth]{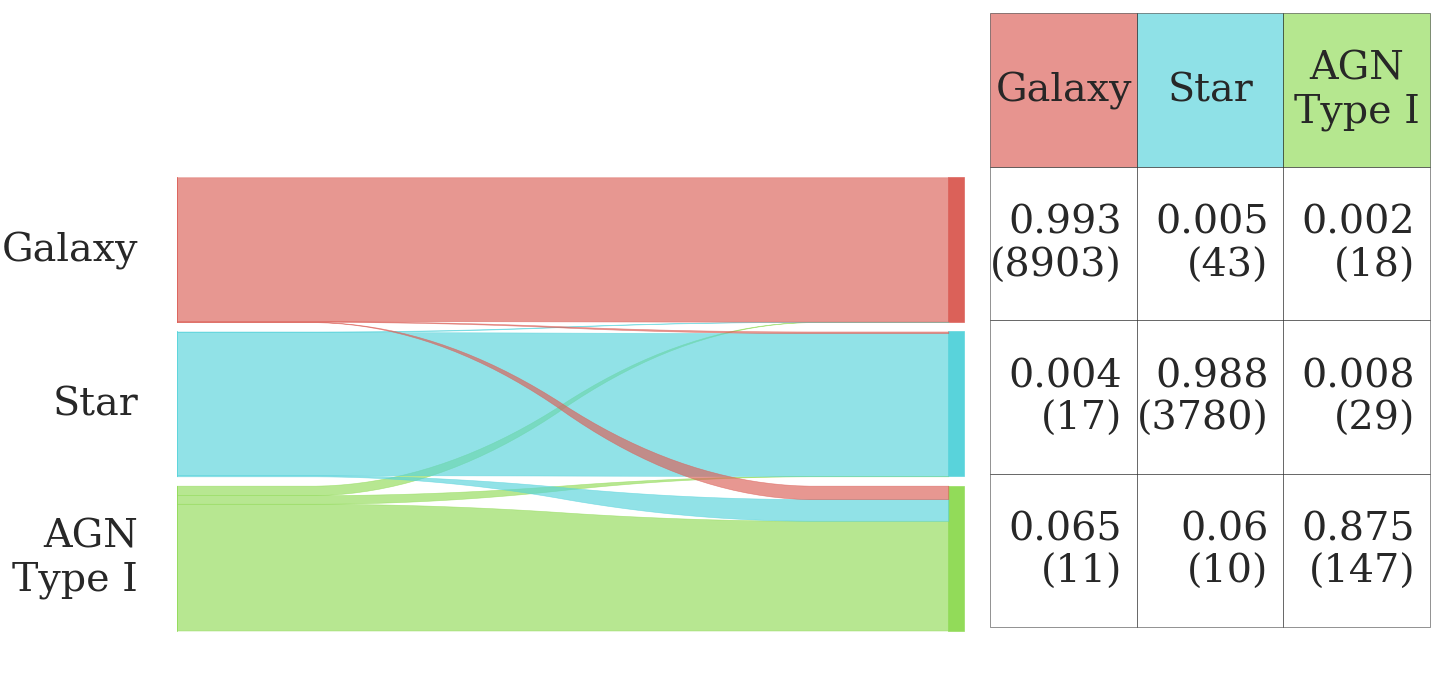}
\includegraphics[width=0.46\textwidth]{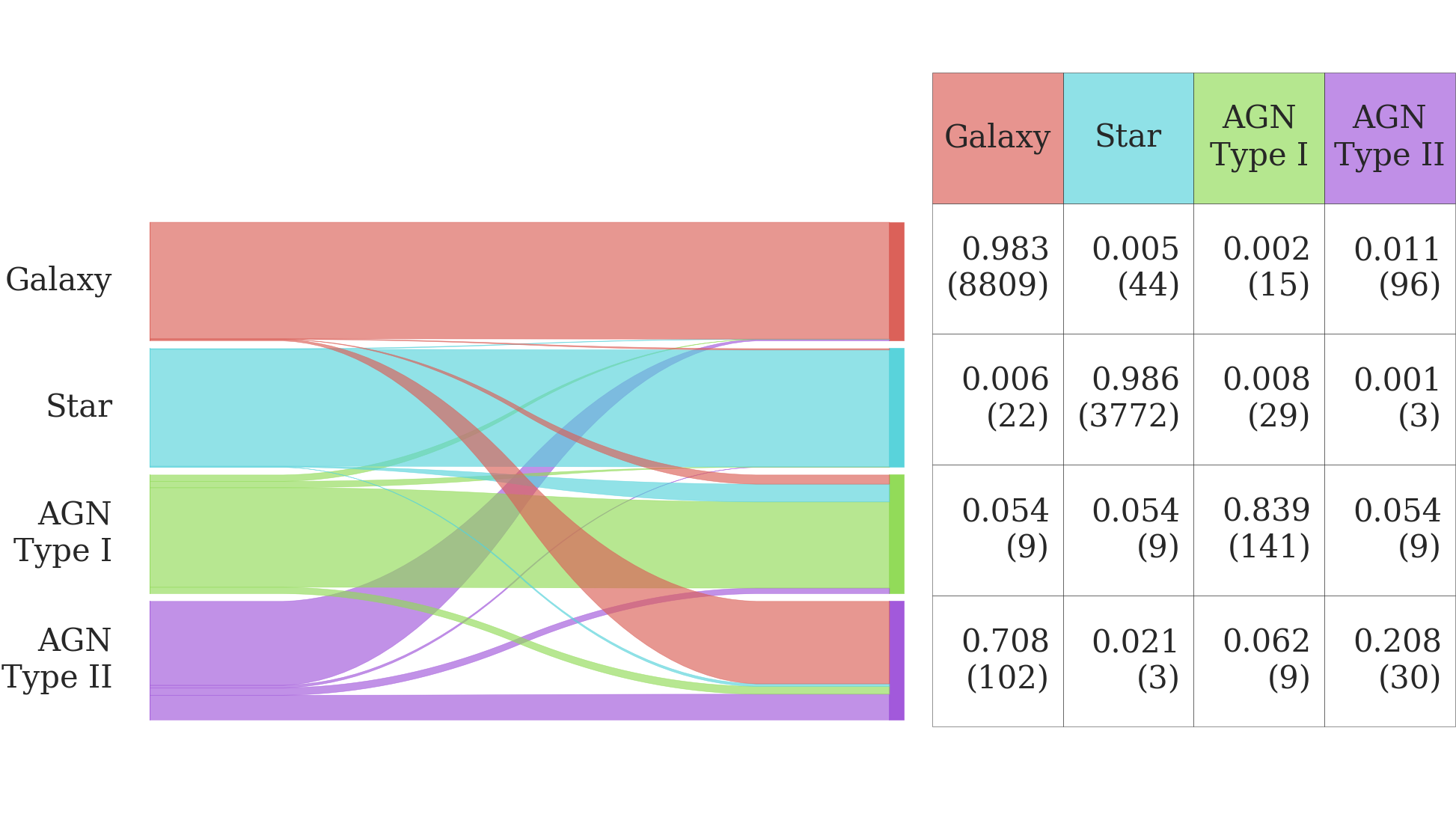}
\par\end{flushright}
\caption{Normalized Sankey diagram and confusion matrices for the 2,3 and 4-class models. The width of the left side of each band in the Sankey diagram corresponds to the fraction of objects belonging to each class that the model predicts as belonging to each of the output classes. The width of the right side of each strip represents the fraction of objects predicted to belong to each class made up by objects truly belonging to each class. In the confusion matrices, each row represents the true class of objects and each column represents the predicted class. The fraction of objects and the absolute number (in brackets) of objects in each true/predicted bin is shown. The 2-class model is evaluated on the $i_{\mathrm{AB}}<25$ sample, while the 3 and 4-class models are tested on the $i_{\mathrm{AB}}<21$ sample.
\label{fig:agnCM}}
\end{figure}

Additionally, we also train a 3-class model which considers only stars, galaxies, and Type I AGN. Figure \ref{fig:agnCM} shows the results of the method applied to the 4-class and 3-class cases.

The four-class model classifies only 21\% of Type II AGN correctly. In fact, the majority of objects predicted to be Type II AGN are galaxies.  This is likely because Type II AGN appear extended like galaxies, but none of the optical colour features are useful in distinguishing them from the galaxy population.  The optical observations available to us in the CLAUDS+HSC-SSP dataset are simply insufficient to identify Type II AGN, and their inclusion in the classification model worsens performance for all other classes. Adding observations where Type II AGN have distinctive characteristics to the input feature list can be expected to make the model better able to recognize these objects. For example, adding photometry at mid-infrared wavelengths \citep{irAGN}, where the colours of Type II AGN continua are redder than those of Type I AGN or galaxies, should boost performance. In this context, various studies have demonstrated that AGN can be classified on the basis of their IR colours (e.g. \citealt{irAGN}, \citealt{wiseAgn}, \citealt{agnIRcolor}.)  and future work might focus on incorporating IR observations into the {\tt XGBoost} model.

Type I AGN, on the other hand, appear morphologically  much like stars but \emph{can} be distinguished using their optical colours. When we eliminate Type II AGN from the classification and train a model  (the 3-class model above) to classify stars, galaxies and Type I AGN, the model performs well: it correctly classifies 87.5\% of Type I AGN without introducing a significant misclassification rate for stars since 98.8\% of stars and 99.3\% of galaxies are classified correctly. As with the four-class model, these numbers could be improved with the addition of further data, such as IR photometry. We caution the reader that at present AGN classifications made with this method should not be used for research applications.

As with the binary classification of stars and galaxies discussed in Section \ref{sec:stargalgen}, the relatively shallow depth of this training sample means that the redshift range of type I AGN represented in the catalog is limited. The model's performance will worsen as the depth of the sample it is applied to increases beyond the flux limit of the training data. A deeper training sample can be expected to improve the model's performance as it will provide a more representative range of colours for type I AGN at higher redshifts that are most probable for fainter objects.


\section{Summary \& Conclusions}

We have described our procedure for classifying stars, galaxies and (experimentally) AGN in the CLAUDS+HSC-SSP $u^*grizy$ catalog using GBT. To predict the class of each object, we first apply an ensemble of {\tt XGBoost}  models, which we call the Stage I Model, to identify a high-purity sample of stars which are used to build a PSF map across the catalog area in each band. This PSF map is used to homogenize all seeing-dependent parameters in the catalog. The homogenized parameters, along with object colours, are then fed into another ensemble of {\tt XGBoost} (the Stage II Model), to provide our final classification.

An analysis of feature importance demonstrates that in addition to colours, {\tt FLUX\_RADIUS} measurements and  $\mu_\mathrm{max}-\mathrm{mag_{Kron}}$ in every band are informative to the classification.Our binary star/galaxy classification model performs better than SExtractor's {\tt CLASS\_STAR} in our best-seeing band ($i$) and produces well-calibrated class probabilities. With a threshold set to maximize the weighted F1 score for our sample, the model achieves 99.8\% galaxy completeness at 96.4\% star purity over a flux-limited, $i_\mathrm{AB}<25$, sample in the COSMOS field. This threshold can in principle be modified to optimize the completeness or purity for different science cases. 

The model ensemble that we apply to the entirety of the 18.60~deg$^2$ CLAUDS+HSC catalog is trained using data from the COSMOS field. In most directions on the sky, the number and types of galaxies represented should be relatively consistent, but the density and types of stars in a sample is highly dependent on the line of sight through the Galaxy. 
To test  that such differences do not significantly degrade performance in other areas of the survey, we apply the model trained in the COSMOS field to an area in the SXDS field.  Using the same method to select a sample of true stars in SXDS as we do for the training data, we find that the binary model's performance is only slightly degraded. This slight degradation may reflect an underlying differences between the populations in the two fields. The evaluation metrics we consider which require the selection of a threshold show a more pronounced decrease in performance likely because our threshold selection is sensitive to the specific star/galaxy ratio in the training field.  

In addition to examining the model's performance on data in other areas of the sky, we consider its ability to extrapolate its classification to objects fainter than those in the sample on which it was trained. We find that successful extrapolation depends primarily on the range of galaxy redshifts represented in the training sample. For our $u^*grizy$ catalog, the completeness of the predicted galaxy sample will remain acceptably high when the model is applied to objects $\sim 1-2$ magnitudes fainter than the limit of the training data. 

We then explore the efficacy of our method for classifying AGN in addition to stars and galaxies. For this, we use a complete sample of AGN in the COSMOS field to train multiclass models to identify stars, galaxies and AGN. Our 3-class star/galaxy/Type I AGN classification model is able to identify 87.5\% of Type I AGN at $i_{\mathrm{AB}}<21$ while correctly classifying 99.3\% and 98.8\% of galaxies and stars, respectively. With the requirements of a specific science case in mind, the method used to convert the probabilistic output of the model could be altered to select more pure or complete samples of any individual class. Our 4-class model, which includes Type II AGN, performs poorly, likely  because of the lack of distinctive features of Type II AGN in the $u^*grizy$ observations. Future work could expand on the results presented here, adding IR observations to the list of input features of the {\tt XGBoost} model ensemble, which will allow the model to identify Type II AGN on the basis of their characteristic red colours.

The GBT framework for object classification in the CLAUDS+HSC-SSP dataset that we presented here provides a potentially flexible and robust method for labelling objects in catalog data.  Our classifications have already been used in constructing UV galaxy luminosity functions from CLAUDS+HSC data \citep{claudsUVLF} and are being used in several other ongoing studies based on this dataset.

\section*{Acknowledgements}

The observations presented here were performed with care and respect from the summit of Maunakea which is a significant cultural and historic site.  This work is based on observations obtained with MegaPrime/MegaCam, a joint project of CFHT and CEA/DAPNIA, at the Canada-France-Hawaii Telescope (CFHT) which is operated by the National Research Council (NRC) of Canada, the Institut National des Sciences de l'Univers of the Centre National de la Recherche Scientifique (CNRS) of France, and the University of Hawaii. This research uses data obtained through the Telescope Access Program (TAP), which has been funded by the National Astronomical Observatories, Chinese Academy of Sciences, and the Special Fund for Astronomy from the Ministry of Finance. This work uses data products from TERAPIX and the Canadian Astronomy Data Centre. It was carried out using resources from Compute Canada and Canadian Advanced Network For Astrophysical Research (CANFAR) infrastructure. We are grateful for research support funding from the Natural Sciences and Engineering Research Council (NSERC) of Canada.

This work is also based on data collected at the Subaru Telescope and retrieved from the HSC data archive system, which is operated by the Subaru Telescope and Astronomy Data Center at National Astronomical Observatory of Japan. The Hyper Suprime-Cam (HSC) collaboration includes the astronomical communities of Japan and Taiwan, and Princeton University. The HSC instrumen-
tation and software were developed by the National Astronomical Observatory of Japan (NAOJ), the Kavli Institute for the Physics and Mathematics of the Universe (Kavli IPMU), the University of Tokyo, the High Energy Accelerator Research Organization (KEK), the Academia Sinica Institute for Astronomy and Astrophysics in Taiwan (ASIAA), and Princeton University. Funding was contributed by the FIRST program from Japanese Cabinet Office, the Ministry of Education, Culture, Sports, Science and Technology (MEXT), the Japan Society for the Promotion of Science (JSPS), Japan Science and Technology Agency (JST), the Toray Science Foundation, NAOJ, Kavli IPMU, KEK, ASIAA, and Princeton University. This paper makes use of software developed for the Large Synoptic Survey Telescope. We thank the LSST Project for making their code available as free software at http://dm.lsst.org.

This work made use of AstroPy, a community-developed core Python package for Astronomy \citep{astropy}, {\tt scikit-learn} \citep{sklearn}, AstroML \citep{astroml}, NumPy, SciPy \citep{scipy}, and MatPlotLib \citep{maplotlib}.

\section*{Data availability}
Hyper Suprime-Cam Subaru Strategic Program Public Data are available at https://hsc-release.mtk.nao.ac.jp, while CLAUDS data are expected to be accessible by the end of 2021 from 
https://www.ap.smu.ca/\textasciitilde sawicki/sawicki/CLAUDS.html.




\bibliographystyle{mnras}
\bibliography{references} 



\appendix


\bsp	
\label{lastpage}
\end{document}